\documentclass[aps,rmp,twocolumn,groupedaddress,pdfLaTeX]{revtex4-2}

\setcitestyle{square,comma,numbers}

\usepackage[dvips]{graphicx}
\usepackage[usenames,svgnames]{xcolor}
\usepackage{color}
\usepackage{url}
\usepackage{natbib}
\usepackage[colorlinks=true,linkcolor=magenta,citecolor=magenta,breaklinks=true]{hyperref}
\usepackage{amsmath,amssymb,amsfonts,bm}

\begin{document}


\title{Shortcuts to adiabaticity: theoretical framework, relations between different methods, and versatile approximations}


\author{Takuya Hatomura}
\email[]{takuya.hatomura@ntt.com}
\affiliation{NTT Basic Research Laboratories \& NTT Research Center for Theoretical Quantum Information, NTT Corporation, Kanagawa 243-0198, Japan}


\date{\today}

\begin{abstract}
Shortcuts to adiabaticity guide given systems to final destinations of adiabatic control via fast tracks. 
Various methods were proposed as varieties of shortcuts to adiabaticity. 
Basic theory of shortcuts to adiabaticity was established in the 2010s, but it has still been developing and many fundamental findings have been reported. 
In this Topical Review, we give a pedagogical introduction to theory of shortcuts to adiabaticity and revisit relations between different methods. 
Some versatile approximations in counterdiabatic driving, which is one of the methods of shortcuts to adiabaticity, will be explained in detail. 
We also summarize recent progress in studies of shortcuts to adiabaticity. 
\end{abstract}


\maketitle


%
%
\section{\label{Sec.intro}Introduction}

Quantum information science is not mere theory any longer, but it has potential to change the real world in these noisy intermediate-scale quantum (NISQ)~\cite{Preskill2018} and post-NISQ eras beyond academic interests. 
Indeed, recent rapid growth of this field could be regarded as second quantum revolution which is a transition from science to technology~\cite{Deutsch2020}. 
The realm of quantum technologies covers a wide range of applications, e.g., computation, simulation, comunication, sensing, etc~\cite{Acin2018}.

Quantum technologies have ability to outperform corresponding classical technologies because of their quantum nature~\cite{Preskill2018,Deutsch2020,Acin2018}. 
However, quantum advantage is easily spoiled by small deviations from ideal implementation. 
Coherent errors may be suppressed by development of technologies, but decoherence is an inevitable phenomenon in quantum systems and it imposes severe restriction on a runtime of quantum processing. 
Precise and fast quantum control is a cornerstone of quantum technologies.

Adiabatic control is one of the basic ways to precisely control given systems and various applications of it have been proposed~\cite{Gaubatz1990,Kadowaki1998,Farhi2000,Roland2002}. 
A guiding principle of adiabatic control is the adiabatic theorem~\cite{Kato1950}, which enables us to track energy eigenstates. 
However, it requires a long operation time to satisfy the adiabatic condition~\cite{Albash2018}. 
It means that adiabatic control suffers from bad influence of decoherence for a long time. 
Moreover, algorithms with long runtimes are not useful.

Many efforts have been made to reduce operation times of quantum adiabatic algorithms~\cite{Unanyan1997,Emmanouilidou2000,Steffen2003,Shapiro2007}.
Shortcuts to adiabaticity were proposed as systematic ways to overcome this drawback of adiabatic control~\cite{Demirplak2003,Demirplak2008,Berry2009,Chen2010,Masuda2008,Masuda2010}. 
There are various methods in shortcuts to adiabaticity, which enable us to realize speedup of adiabatic time evolution or to obtain final results of adiabatic control without requiring the adiabatic condition. 
In this Topical Review, we focus on three basic methods of shortcuts to adiabaticity, i.e., counterdiabatic driving~\cite{Demirplak2003,Demirplak2008,Berry2009}, invariant-based inverse engineering~\cite{Chen2010}, and fast-forward scaling~\cite{Masuda2008,Masuda2010}. 
In particular, various versatile approximations have been proposed for counterdiabatic driving.

Counterdiabatic driving is a method to obtain adiabatic time evolution within a short time~\cite{Demirplak2003,Demirplak2008,Berry2009}. 
In this method, we introduce additional driving, which counteracts diabatic changes, and realize the adiabatic state. 
The additional term is called the counterdiabatic term or the counterdiabatic Hamiltonian. 
In the early stage, exact application of counterdiabatic driving was limited to systems which have small degrees of freedom or specific properties~\cite{Muga2010,Chen2010a,delCampo2012,delCampo2013,Jarzynski2013,Takahashi2013,Deffner2014,Okuyama2016,Hatomura2018b}. 
Even approximate application required approximate solutions of eigenstate problems (see, Ref.~\cite{Torrontegui2013,Guery-Odelin2019} and references therein). 
Numerical optimization of approximate counterdiabatic terms with ansatzes on operator forms was discussed~\cite{Saberi2014,Campbell2015}, but it still required knowledge of target dynamics.

A breakthrough was a variational approach~\cite{Sels2017} to construction of counterdiabatic Hamiltonians. 
In this approach, a trial counterdiabatic Hamiltonian is introduced and its time dependence is determined by using a condition for the counterdiabatic Hamiltonian~\cite{Jarzynski2013}  as a cost function. 
A natural way to determine operator forms of trial counterdiabatic Hamiltonians was also proposed~\cite{Claeys2019}. 
Remarkably, this approach does not require knowledge of energy eigenstates.

Invariant-based inverse engineering is a method to obtain a final state of adiabatic time evolution within a short time~\cite{Chen2010}. 
In this method, we introduce target dynamics, which is identical with the adiabatic state at the initial time and the final time, and inversely specify a Hamiltonian. 
Note that target dynamics is not necessarily equal to the adiabatic state at intermediate time. 
By properly choosing target dynamics, we can avoid introduction of unwanted driving terms. 
Similarly to counterdiabatic driving, invariant-based inverse engineering was first applied to simple or special systems~\cite{Torrontegui2011,Chen2012,Ruschhaupt2012,Gungordu2012}. 
Moreover, relation between invariant-based inverse engineering and counterdiabatic driving was discussed and similar mathematical structure was pointed out~\cite{Chen2011,Takahashi2017a}.

Fast-forward scaling is a method to obtain target dynamics within a different time~\cite{Masuda2008}. 
In this method, we introduce additional driving which enables us to change timescale of given dynamics with additional phase factors. 
Fast forward scaling can also be applied to speedup of adiabatic time evolution~\cite{Masuda2010}. 
Various applications were proposed~\cite{Masuda2011,Masuda2012,Masuda2014,Deffner2015,Jarzynski2017,Patra2017,Setiawan2019}. 
Moreover, relations between fast-forward scaling of adiabatic time evolution and other methods of shortcuts to adiabaticity have been discussed~\cite{Torrontegui2012,Jarzynski2017,Patra2017,Setiawan2017}. 
Application of fast-forward scaling to other methods of shortcuts to adiabaticity was also discussed~\cite{Takahashi2014}.

There are comprehensive systematic reviews on shortcuts to adiabaticity~\cite{Torrontegui2013,Guery-Odelin2019}. 
Moreover, other reviews can also be found. 
In Ref.~\cite{DelCampo2015}, application of counterdiabatic driving to critical many-body systems was mentioned. 
Application of counterdiabatic driving and fast-forward scaling to atomic, molecular, and chemical physics was summarized in a book chapter~\cite{Masuda2016}. 
In Ref.~\cite{Kolodrubetz2017}, counterdiabatic driving was discussed from the viewpoint of geometry and nonadiabatic response in quantum and classical systems. 
Reference~\cite{delCampo2019} is an editorial for a special issue focusing on shortcuts to adiabaticity. 
In the first half of Ref.~\cite{Takahashi2019}, a short review on shortcuts to adiabaticity was given. 
We can also find a review focusing on fast-forward scaling~\cite{Masuda2022}. 
Pedagogical introductions to counterdiabatic driving~\cite{Nakahara2022} and to invariant-based inverse engineering~\cite{Takahashi2022} are also available. 
However, theory of shortcuts to adiabaticity is a rapidly growing research area. 
Even after publication of the comprehensive systematic review in 2019~\cite{Guery-Odelin2019}, a lot of great progress has been reported. 
One of the purposes of this Topical Review is to introduce these recent results.

We also provide pedagogical overviews on background and theory of shortcuts to adiabaticity. 
In Sec.~\ref{Sec.background}, we explain minimal knowledge for understanding shortcuts to adiabaticity. 
Three methods of shortcuts to adiabaticity are explained in Sec.~\ref{Sec.theory.STA}. 
In particular, a simple formulation of fast-forward scaling is given. 
Then, we discuss relations between different methods of shortcuts to adiabaticity in Sec.~\ref{Sec.rel.STA}. 
Relations between fast-forward scaling and other methods of shortcuts to adiabaticity are revisited by using the simple formulation of fast-forward scaling. 
Versatile approximations in counterdiabatic driving and a way to evaluate performance of approximate shortcuts to adiabaticity are explained in Sec.~\ref{Sec.ver.app}. 
Mainly, we provide a unified viewpoint of versatile approximations in counterdiabatic driving. 
Finally, other recent progress is summarized in Sec.~\ref{Sec.progress}. 
Note that we may use different characters from conventional notations to avoid duplicate use of them and confusion.

%
%
\section{\label{Sec.background}Theoretical background}

Theoretical background of shortcuts to adiabaticity is explained below. 
In this Topical Review, we mainly focus on closed quantum systems. 
Namely, dynamics is governed by the Schr\"odinger equation
\begin{equation}
i\hbar\frac{\partial}{\partial t}|\Psi(t)\rangle=\hat{H}(t)|\Psi(t)\rangle, 
\label{Eq.Seq}
\end{equation}
where $\hat{H}(t)$ is a Hamiltonian of a given system and $|\Psi(t)\rangle$ is its dynamics. 
Note that we shortly mention recent progress on shortcuts to adiabaticity for open quantum systems in Sec.~\ref{Sec.open}.

%
%
\subsection{\label{Sec.nad}Nonadiabatic transitions}

In this section, we consider origin of nonadiabatic transitions. 
Suppose that the Hamiltonian $\hat{H}(t)$ and the dynamics $|\Psi(t)\rangle$ are spanned by a laboratory basis. 
We introduce a unitary transformation $\hat{V}(t)$ which diagonalizes the Hamiltonian $\hat{H}(t)$, i.e., $\hat{V}^\dag(t)\hat{H}(t)\hat{V}(t)=\mathrm{diag}\bm{(}E_1(t),E_2(t),\ldots\bm{)}$. 
Here, $E_n(t)$ ($n=1,2,\ldots$) is the eigenenergy, or in other words, the energy level, and we express its eigenstate as $|n(t)\rangle$. 
Note that the $(m,n)$ element of the matrix $\hat{V}(t)$ is given by the $m$th element of the eigenstate $|n(t)\rangle$. 
Then, the Schr\"odinger equation (\ref{Eq.Seq}) can be rewritten as
\begin{equation}
i\hbar\frac{\partial}{\partial t}|\tilde{\Psi}(t)\rangle=[\hat{V}^\dag(t)\hat{H}(t)\hat{V}(t)-i\hbar\hat{V}^\dag(t)\bm{(}\partial_t\hat{V}(t)\bm{)}]|\tilde{\Psi}(t)\rangle,
\label{Eq.nad2}
\end{equation}
where $|\tilde{\Psi}(t)\rangle=\hat{V}^\dag(t)|\Psi(t)\rangle$ and $\partial_t=\partial/\partial t$ (see, e.g., Ref.~\cite{Demirplak2003}). 
Since the first term is diagonal in the energy-eigenstate basis (the adiabatic basis), the off-diagonal elements of the second term cause transitions between different energy levels, i.e., nonadiabatic transitions.

Equivalently, by expanding the dynamics in the adiabatic basis, $|\Psi(t)\rangle=\sum_nc_n(t)e^{-\frac{i}{\hbar}\int_0^tdt^\prime E_n(t^\prime)}|n(t)\rangle$, where 
$c_n(t)$ is a time-dependent coefficient, we obtain equation of motion for the coefficient
\begin{equation}
i\hbar\frac{\partial}{\partial t}c_n(t)+i\hbar\sum_m\langle n(t)|\partial_t m(t)\rangle c_m(t)=0, 
\label{Eq.nad}
\end{equation}
from the Schr\"odinger equation (\ref{Eq.Seq}). 
The amount of nonadiabatic transitions is characterized by time evolution of the distribution $\{|c_n(t)|^2\}_{n=1,2,\ldots}$. 
Note that the second term in Eq.~(\ref{Eq.nad}) corresponds with that in Eq.~(\ref{Eq.nad2}).

%
%
\subsection{Adiabatic time evolution}

In this section, we introduce the adiabatic condition and the adiabatic state. 
When the second term of Eq.~(\ref{Eq.nad2}) is much smaller than the first term, nonadiabatic transitions are suppressed. 
By focusing on certain two energy levels, it can be formulated as the adiabatic condition
\begin{equation}
\hbar\left|\frac{\langle n(t)|\partial_tm(t)\rangle}{E_m(t)-E_n(t)}\right|=\hbar\left|\frac{\langle n(t)|\bm{(}\partial_t\hat{H}(t)\bm{)}|m(t)\rangle}{\bm{(}E_m(t)-E_n(t)\bm{)}^2}\right|\ll1, 
\label{Eq.adcond}
\end{equation}
where the numerator $\langle n(t)|\partial_tm(t)\rangle$ is the off-diagonal element of the second term and the denominator $\bm{(}E_m(t)-E_n(t)\bm{)}$ is the amplitude of the corresponding elements in the first term (see, e.g., Ref.~\cite{Demirplak2003}). 
The adiabatic condition (\ref{Eq.adcond}) claims that nonadiabatic transitions do not take place when the Hamiltonian changes slowly in time against the energy gap. 
This statement is also known as the adiabatic theorem~\cite{Kato1950}. 
For rigorous discussion on the adiabatic theorem and the adiabatic condition, see Ref.~\cite{Albash2018}.

When the adiabatic condition (\ref{Eq.adcond}) is satisfied, the off-diagonal elements of the second term in Eq.~(\ref{Eq.nad2}) [Eq.~(\ref{Eq.nad})] can be regarded as perturbation terms. 
By neglecting these perturbation terms, we can formally solve Eq.~(\ref{Eq.nad}), and then we obtain the adiabatic state~\cite{Berry1984}
\begin{equation}
\begin{aligned}
|\Psi_\mathrm{ad}(t)\rangle=\sum_n&c_n(0)e^{-\frac{i}{\hbar}\int_0^tdt^\prime E_n(t^\prime)} \\
&\times e^{-\int_0^tdt^\prime\langle n(t^\prime)|\partial_{t^\prime}n(t^\prime)\rangle}|n(t)\rangle, 
\end{aligned}
\label{Eq.ad}
\end{equation}
as the unperturbed solution of Eq.~(\ref{Eq.Seq}). 
Here, the first phase factor is known as the dynamical phase and the second one is known as the geometric phase.

A geometric property of the second phase factor can be confirmed by explicitly writing parameter dependence of the energy eigenstate, i.e., $|n(t)\rangle\to|n(\bm{\lambda})\rangle$, where $\bm{\lambda}=\bm{\lambda}(t)=\bm{(}\lambda_1(t),\lambda_2(t),\ldots\bm{)}$ is a vector of time-dependent parameters. 
Then, the second phase factor is expressed by the line integral in parameter space $e^{-\int_{\mathcal{C}}d\bm{\lambda}\cdot\langle n(\bm{\lambda})|\bm{\nabla}_{\bm{\lambda}}n(\bm{\lambda})\rangle}$, where $\bm{\nabla}_{\bm{\lambda}}=\partial/\partial\bm{\lambda}$ and $\mathcal{C}$ is a path in parameter space. 
Namely, this phase does not depend on time schedule of parameters, but it depends on the path in parameter space. 
When the path $\mathcal{C}$ is a closed loop, it gives the Berry phase~\cite{Berry1984}.

%
%
\subsection{\label{Sec.LR}Lewis-Riesenfeld theory}

In this section, we introduce the Lewis-Riesenfeld theory~\cite{Lewis1969}, in which arbitrary dynamics is expressed like the adiabatic state (\ref{Eq.ad}). 
The dynamical invariant $\hat{F}(t)$, also known as the Lewis-Riesenfeld invariant, is an Hermitian operator which satisfies the von Neumann equation
\begin{equation}
i\hbar\frac{\partial}{\partial t}\hat{F}(t)-\mathcal{L}_H\hat{F}(t)=0, 
\label{Eq.dyinv}
\end{equation}
where $\mathcal{L}_H$ is the Liouvillian superoperator of closed quantum mechanics, $\mathcal{L}_H\bullet=[\hat{H}(t),\bullet]$. 
Note that the density operator is a trivial example of the dynamical invariant. 
Each eigenvalue of the operator $\hat{F}(t)$ is conserved during time evolution, whereas the Hamiltonian depends on time, and thus it is called the dynamical invariant~\cite{Lewis1969}. 
It is also notable that the left-hand side of Eq.~(\ref{Eq.dyinv}) in the classical limit can be regarded as the total derivative of a classical quantity (see, e.g., Ref.~\cite{Takahashi2022}).

It was shown that the solution of the Schr\"odinger equation (\ref{Eq.Seq}) can be expressed as
\begin{equation}
|\Psi(t)\rangle=\sum_n\tilde{c}_n(0)e^{i\alpha_n(t)}|\phi_n(t)\rangle, 
\label{Eq.sol.Seq}
\end{equation}
by using the eigenvectors of the Lewis-Riesenfeld invariant $\{|\phi_n(t)\rangle\}_{n=1,2,\ldots}$~\cite{Lewis1969}. 
Here, $\tilde{c}_n(0)$ ($n=1,2,\ldots$) is a time-independent coefficient and $\alpha_n(t)$ is given by
\begin{equation}
\alpha_n(t)=\frac{1}{\hbar}\int_0^tdt^\prime\langle\phi_n(t^\prime)|\bm{(}i\hbar\partial_{t^\prime}-\hat{H}(t^\prime)\bm{)}|\phi_n(t^\prime)\rangle,
\label{Eq.LRphase}
\end{equation}
which is called the Lewis-Riesenfeld phase. 
We can easily confirm this fact by considering time derivative of this state (\ref{Eq.sol.Seq}) and by using the following equation
\begin{equation}
\langle\phi_m(t)|\hat{H}(t)|\phi_n(t)\rangle=i\hbar\langle\phi_m(t)|\partial_t\phi_n(t)\rangle,\quad m\neq n,
\label{Eq.offele}
\end{equation}
which can be derived from Eq.~(\ref{Eq.dyinv}).

%
%
\subsection{\label{Sec.inv.eng}Inverse engineering of Hamiltonians}

In this section, we consider inverse engineering of Hamiltonians for realizing given target dynamics. 
There are many approaches to inverse engineering. 
Here we mention two of them.

\begin{itemize}
\item Approach 1. In the first approach, we introduce known target dynamics $|\Psi(t)\rangle$ and a Hamiltonian $\hat{H}(t)$ whose time dependence is undetermined. 
By considering the Schr\"odinger equation (\ref{Eq.Seq}) in a certain basis, i.e., the Schr\"odinger equation for $\Psi_\sigma(t)=\langle\sigma|\Psi(t)\rangle$ with a basis $\{|\sigma\rangle\}$, we can obtain equations for time dependence of the Hamiltonian. 
Note that the basis $\{|\sigma\rangle\}$ may depend on time.

\item Approach 2. In the second approach, we express target dynamics $|\Psi(t)\rangle$ as $|\Psi(t)\rangle=\hat{U}(t)|\Psi(0)\rangle$ by using a unitary operator $\hat{U}(t)$. 
By inversely solving the Schr\"odinger equation (\ref{Eq.Seq}), we find a Hamiltonian
\begin{equation}
\hat{H}(t)=i\hbar\bm{(}\partial_t\hat{U}(t)\bm{)}\hat{U}^\dag(t)
\end{equation}
which generates the target dynamics. 
\end{itemize}

%
%
\section{\label{Sec.theory.STA}Methods of shortcuts to adiabaticity}

Here, we introduce three methods of shortcuts to adiabaticity, i.e., counterdiabatic driving~\cite{Demirplak2003,Demirplak2008,Berry2009}, invariant-based inverse engineering~\cite{Chen2010}, and fast-forward scaling~\cite{Masuda2008,Masuda2010}. 
We show that these methods can be formulated by using inverse engineering approaches in Sec.~\ref{Sec.inv.eng}.

%
%
\subsection{Counterdiabatic driving}

First, we discuss counterdiabatic driving~\cite{Demirplak2003,Demirplak2008,Berry2009}. 
In this method, we introduce additional driving, which counteracts diabatic changes, and the adiabatic state (\ref{Eq.ad}) is realized within a short time.

%
%
\subsubsection{Counterdiabatic terms}

As mentioned in Sec.~\ref{Sec.nad}, nonadiabatic transitions are caused by the off-diagonal elements of the second term in Eq.~(\ref{Eq.nad2}). 
In counterdiabatic driving, we introduce additional driving $\hat{H}_\mathrm{cd}(t)$ as $\hat{H}(t)\to\hat{H}(t)+\hat{H}_\mathrm{cd}(t)$, which counteracts these diabatic terms, i.e., $\hat{V}^\dag(t)\hat{H}_\mathrm{cd}(t)\hat{V}(t)=i\hbar\hat{V}^\dag(t)\bm{(}\partial_t\hat{V}(t)\bm{)}$ except for the diagonal elements~\cite{Demirplak2003}. 
By using the adiabatic basis, it is given by
\begin{equation}
\hat{H}_\mathrm{cd}(t)=i\hbar\sum_{\stackrel{m,n}{(m\neq n)}}|n(t)\rangle\langle n(t)|\partial_tm(t)\rangle\langle m(t)|, 
\label{Eq.cdham}
\end{equation}
which is known as the counterdiabatic Hamiltonian or counterdiabatic terms. 
Here, we assume that the diagonal elements of the counterdiabatic Hamiltonian, which affect phase factors, are zeros. 
Then, we realize the adiabatic state of the original reference Hamiltonian (\ref{Eq.ad}) as the exact solution of the Schr\"odinger equation
\begin{equation}
i\hbar\frac{\partial}{\partial t}|\Psi_\mathrm{ad}(t)\rangle=\hat{H}_\mathrm{ad}(t)|\Psi_\mathrm{ad}(t)\rangle, 
\label{Eq.Seq.ad}
\end{equation}
where
\begin{equation}
\hat{H}_\mathrm{ad}(t)=\hat{H}(t)+\hat{H}_\mathrm{cd}(t), 
\label{Eq.adham} 
\end{equation}
for any time dependence of the reference Hamiltonian $\hat{H}(t)$.

Note that we can rewrite the counterdiabatic Hamiltonian (\ref{Eq.cdham}) as
\begin{equation}
\hat{H}_\mathrm{cd}(t)=\frac{d\bm{\lambda}}{dt}\cdot\hat{\mathcal{A}}(\bm{\lambda}),
\label{Eq.cdham.AGP}
\end{equation}
where
\begin{equation}
\hat{\mathcal{A}}(\bm{\lambda})=i\hbar\sum_{\substack{m,n \\ (m\neq n)}}|n(\bm{\lambda})\rangle\langle n(\bm{\lambda})|\bm{\nabla}_{\bm{\lambda}}m(\bm{\lambda})\rangle\langle m(\bm{\lambda})|,
\label{Eq.AGP}
\end{equation}
by explicitly writing parameter dependence. 
Namely, we can decouple a time-dependent part $d\bm{\lambda}/dt$ from a parameter-dependent part $\hat{\mathcal{A}}(\bm{\lambda})$. 
Here, $\hat{\mathcal{A}}(\bm{\lambda})$ is known as adiabatic gauge potential, which is a generator of energy eigenstates in parameter space. 
For more information, see, e.g., Ref.~\cite{Kolodrubetz2017}.

%
%
\subsubsection{\label{Sec.gene.CD}Inverse engineering of counterdiabatic terms}

Counterdiabatic driving can also be formulated as inverse engineering of Hamiltonians~\cite{Demirplak2008,Berry2009}. 
We consider a unitary operator $\hat{U}_\mathrm{ad}(t)$ which generates the adiabatic state of the reference Hamiltonian (\ref{Eq.ad}), i.e., $|\Psi_\mathrm{ad}(t)\rangle=\hat{U}_\mathrm{ad}(t)|\Psi_\mathrm{ad}(0)\rangle$, and apply the second approach explained in Sec.~\ref{Sec.inv.eng}. 
The explicit expression of this unitary operator is given by
\begin{equation}
\begin{aligned}
\hat{U}_\mathrm{ad}(t)=\sum_n&e^{-\frac{i}{\hbar}\int_0^tdt^\prime E_n(t^\prime)}e^{-\int_0^tdt^\prime\langle n(t^\prime)|\partial_{t^\prime}n(t^\prime)\rangle} \\
&\times|n(t)\rangle\langle n(0)|. 
\end{aligned}
\end{equation}
By formally solving the Schr\"odinger equation (\ref{Eq.Seq.ad}), we obtain
\begin{equation}
\hat{H}_\mathrm{ad}(t)=i\hbar\bm{(}\partial_t\hat{U}_\mathrm{ad}(t)\bm{)}\hat{U}_\mathrm{ad}^\dag(t). 
\end{equation}
We find that the time derivative of the dynamical phase generates the reference Hamiltonian $\hat{H}(t)$, and these of the geometric phase and the energy eigenstate basis $|n(t)\rangle$ constitute the counterdiabatic Hamiltonian (\ref{Eq.cdham}). 
Accordingly, we reproduce the total Hamiltonian (\ref{Eq.adham})~\cite{Demirplak2008,Berry2009}.

%
%
\subsection{Invariant-based inverse engineering}

Next, we discuss invariant-based inverse engineering~\cite{Chen2010}. 
In this method, we consider inverse engineering of a Hamiltonian which generates target dynamics. 
By using the knowledge of the dynamical invariant, we can avoid introduction of unwanted drivng terms.

%
%
\subsubsection{Dynamical modes and inverse engineering}

First, we consider dynamics (\ref{Eq.sol.Seq}) and inverse engineering of a Hamiltonian. 
As mentioned in Sec.~\ref{Sec.LR}, any dynamics can be expressed as Eq.~(\ref{Eq.sol.Seq}). 
The time-evolution operator of dynamics (\ref{Eq.sol.Seq}) is given by
\begin{equation}
\hat{U}(t)=\sum_ne^{i\alpha_n(t)}|\phi_n(t)\rangle\langle\phi_n(0)|, 
\end{equation}
and inverse engineering of a Hamiltonian gives
\begin{equation}
\begin{aligned}
\hat{H}(t)=&-\hbar\sum_n\frac{d\alpha_n(t)}{dt}|\phi_n(t)\rangle\langle\phi_n(t)|\\
&+i\hbar\sum_n|\partial_t\phi_n(t)\rangle\langle\phi_n(t)|, 
\end{aligned}
\label{Eq.inv.ham}
\end{equation}
where we consider the second approach of inverse engineering in Sec.~\ref{Sec.inv.eng}~\cite{Chen2011}. 
Note that the diagonal and off-diagonal elements of the Hamiltonian (\ref{Eq.inv.ham}) give the Lewis-Riesenfeld phase (\ref{Eq.LRphase}) and Eq.~(\ref{Eq.offele}), respectively. 
We also find that an operator
\begin{equation}
\hat{F}(t)=\sum_n\bar{f}_{\phi_n}|\phi_n(t)\rangle\langle\phi_n(t)|,
\label{Eq.inv}
\end{equation}
with an arbitrary time-independent coefficient $\bar{f}_{\phi_n}$ satisfies Eq.~(\ref{Eq.dyinv}) with the Hamiltonian (\ref{Eq.inv.ham}), and thus it is the dynamical invariant. 
The Hamiltonian (\ref{Eq.inv.ham}) generates dynamics (\ref{Eq.sol.Seq}), but it may have unwanted driving terms.

%
%
\subsubsection{Procedure of invariant-based inverse engineering}

Now, we explain the precedure of invariant-based inverse engineering. 
It was first proposed for a system, of which we know a Hamiltonian and its dynamical invariant~\cite{Chen2010}, but it can be applied to general cases as follows~\cite{Sarandy2011,Gungordu2012,Takahashi2013a,Torrontegui2014}. 
We introduce a set of time-independent operators $\{\hat{X}_k\}_k$, and assume that a Hamiltonian is described by using its subset $A\subset\{\hat{X}_k\}_k$ and its dynamical invariant is described by using another subset $B\subset\{\hat{X}_k\}_k$, i.e., 
\begin{equation}
\hat{H}(t)=\sum_{\substack{k \\ (\hat{X}_k\in A)}}h_k(t)\hat{X}_k,\quad\hat{F}_k(t)=\sum_{\substack{k \\ (\hat{X}_k\in B)}}\tilde{f}_k(t)\hat{X}_k, 
\label{Eq.ham.dyinv.alg}
\end{equation}
where $h_k(t)$ and $\tilde{f}_k(t)$ are time-dependent parameters. 
To satisfy Eq.~(\ref{Eq.dyinv}), these subset $A,B$ must satisfy
\begin{equation}
[\hat{X}_k,\hat{X}_l]\in B,\quad\text{for }\hat{X}_k\in A,\ \hat{X}_l\in B. 
\end{equation}
For given algebraic structure
\begin{equation}
[\hat{X}_j,\hat{X}_k]=i\sum_lT_{jkl}\hat{X}_l,
\end{equation}
with an antisymmetric tensor $T_{jkl}$, we obtain
\begin{equation}
i\hbar\frac{\partial}{\partial t}\tilde{f}_j(t)=\sum_{j,k,l}T_{jkl}h_k(t)\tilde{f}_l(t),
\label{Eq.cond.ham.dyinv}
\end{equation}
This equation can be viewed as linear equations for the Hamiltonian parameters $\{h_k(t)\}_k$.

Now, we introduce target dynamics in the form of Eq.~(\ref{Eq.sol.Seq}) by using eigenvectors of the dynamical invariant in Eq.~(\ref{Eq.ham.dyinv.alg}). 
Then, we obtain parameter schedule of the Hamiltonian in Eq.~(\ref{Eq.ham.dyinv.alg}) which is the solution of Eq.~(\ref{Eq.cond.ham.dyinv}) with the auxiliary equation (\ref{Eq.LRphase}). 
Remarkably, this method does not require additional driving terms since the operator form is priorly given.

This method enables us to obtain the same final state with adiabatic time evolution. 
One may require commutativity between the Hamiltonian and the dynamical invariant at the initial time and the final time to avoid unwanted transitions.

%
%
\subsection{Fast-forward scaling}

Finally, we discuss fast-forward scaling~\cite{Masuda2008}. 
In this method, we inversely construct potential or additional driving which enables us to obtain target dynamics within a different time.

%
%
\subsubsection{Fast-forward potential}

Historically, fast-forward scaling was proposed to speedup dynamics of a single-particle system in Ref.~\cite{Masuda2008}. 
To avoid confusion, we first introduce inverse engineering of local potential for realizing target dynamics of a single-particle system based on a method in Ref.~\cite{Torrontegui2012}, and after that, we consider fast-forward of it. 
Note that it corresponds with the first approach in Sec.~\ref{Sec.inv.eng}. 
We introduce a wave function of a quantum particle $\Psi(\bm{x},t)=\langle\bm{x}|\Psi(t)\rangle$, where $\bm{x}$ is the coordinate representation. 
The Schr\"odinger equation (\ref{Eq.Seq}) for the wave function is given by
\begin{equation}
i\hbar\frac{\partial}{\partial t}\Psi(\bm{x},t)=-\frac{\hbar^2}{2m}\bm{\nabla}^2\Psi(\bm{x},t)+V(\bm{x},t)\Psi(\bm{x},t),
\label{Eq.Seq.coordinate}
\end{equation}
where $\bm{\nabla}=\partial/\partial\bm{x}$ and $V(\bm{x},t)$ is time-dependent potential. 
Since the wave function is a complex number, we can express it as
\begin{equation}
\Psi(\bm{x},t)=r(\bm{x},t)e^{i\theta(\bm{x},t)},\quad r(\bm{x},t),\theta(\bm{x},t)\in\mathbb{R},
\end{equation}
where $r(\bm{x},t)$ is the amplitude and $\theta(\bm{x},t)$ is the phase of the wave function. 
Then, by inversely solving Eq.~(\ref{Eq.Seq.coordinate}), we obtain
\begin{equation}
\mathrm{Re}V(\bm{x},t)=-\hbar\frac{\partial\theta(\bm{x},t)}{\partial t}+\frac{\hbar^2}{2m}\left[\frac{\bm{\nabla}^2r(\bm{x},t)}{r(\bm{x},t)}-\bm{(}\bm{\nabla}\theta(\bm{x},t)\bm{)}^2\right],
\label{Eq.relpote}
\end{equation}
and
\begin{equation}
\begin{aligned}
&\mathrm{Im}V(\bm{x},t)=\\
&\hbar\frac{\partial_tr(\bm{x},t)}{r(\bm{x},t)}+\frac{\hbar^2}{2m}\left[\bm{\nabla}^2\theta(\bm{x},t)+2\frac{\bm{(}\bm{\nabla}\theta(\bm{x},t)\bm{)}\cdot\bm{(}\bm{\nabla}r(\bm{x},t)\bm{)}}{r(\bm{x},t)}\right].
\label{Eq.impote}
\end{aligned}
\end{equation}
We usually require $\mathrm{Im}V(\bm{x},t)=0$ for physical systems.

Since $|\Psi(\bm{x},t)|^2=|r(\bm{x},t)|^2$, we can obtain the same probability distribution with certain target dynamics $\Psi(\bm{x},t)$ by giving appropriate amplitude $r(\bm{x},t)$. 
For such amplitude $r(\bm{x},t)$, we can determine the phase $\theta(\bm{x},t)$ by setting $\mathrm{Im}V(\bm{x},t)=0$ in Eq.~(\ref{Eq.impote}), and then we obtain potential $V(\mathrm{x},t)$ by substituting given $r(\bm{x},t)$ and resulting $\theta(\bm{x},t)$ for Eq.~(\ref{Eq.relpote}).

Now, we discuss fast-forward scaling of this inversely engineered dynamics~\cite{Masuda2008}. 
We consider a fast-forward state
\begin{equation}
\Psi_\mathrm{FF}(\bm{x},t)=\Psi(\bm{x},s)e^{if(\bm{x},t)},
\end{equation}
with rescaled time $s=s(t)$ and additional phase $f(\bm{x},t)$. 
Since $|\Psi_\mathrm{FF}(\bm{x},t)|^2=|\Psi(\bm{x},s)|^2$, we can obtain the same probability distribution with certain target dynamics $\Psi(\bm{x},s)$ at time $s$ within a different time $t$. 
By inversely solving the Schr\"odinger equation
\begin{equation}
i\hbar\frac{\partial}{\partial t}\Psi_\mathrm{FF}(\bm{x},t)=-\frac{\hbar^2}{2m}\bm{\nabla}^2\Psi_\mathrm{FF}(\bm{x},t)+V_\mathrm{FF}(\bm{x},t)\Psi_\mathrm{FF}(\bm{x},t),
\end{equation}
we can find the fast-forward potential $\hat{V}_\mathrm{FF}(\bm{x},t)$. 
The imaginary part of the fast-forward potential becomes
\begin{equation}
\mathrm{Im}V_\mathrm{FF}(\bm{x},t)=\frac{ds}{dt}\mathrm{Im}V(\bm{x},s), 
\end{equation}
when the additional phase is given by $f(\bm{x},t)=(ds/dt-1)\theta(\bm{x},s)$. 
With this phase, the real part of the fast-forward potential is given by
\begin{equation}
\begin{aligned}
\mathrm{Re}V_\mathrm{FF}(\bm{x},t)=&\mathrm{Re}V(\bm{x},s)-\hbar\frac{d^2s}{dt^2}\theta(\bm{x},s)\\
&-\hbar\left[\left(\frac{ds}{dt}\right)^2-1\right]\frac{\partial\theta(\bm{x},s)}{\partial s}\\
&-\frac{\hbar^2}{2m}\left[\left(\frac{ds}{dt}\right)^2-1\right]\bm{(}\bm{\nabla}\theta(\bm{x},s)\bm{)}^2.
\end{aligned}
\label{Eq.FFpote}
\end{equation}
Namely, for given amplitude of target dynamics $r(\bm{x},t)$, we obtain the phase $\theta(\bm{x},t)$ by solving $\mathrm{Im}V(\bm{x},t)=0$ in Eq.~(\ref{Eq.impote}), and then we find the potential (\ref{Eq.relpote}) and the fast-forward potential (\ref{Eq.FFpote}) which enables us to obtain target dynamics with different time scale.

%
%
\subsubsection{\label{Sec.FF.gen}Generator-based approach}

Here, we introduce a different formulation of fast-forward scaling based on the second approach to inverse engineering in Sec.~\ref{Sec.inv.eng}. 
We consider the following fast-forward state
\begin{equation}
|\Psi_\mathrm{FF}(t)\rangle=\hat{U}_f(t)|\Psi(s)\rangle, 
\label{Eq.FFstate}
\end{equation}
where $\hat{U}_f(t)$ is a unitary operator which does not affect measurement outcomes. 
By inversely solving the Schr\"odinger equation
\begin{equation}
i\hbar\frac{\partial}{\partial t}|\Psi_\mathrm{FF}(t)\rangle=\hat{H}_\mathrm{FF}(t)|\Psi_\mathrm{FF}(t)\rangle, 
\end{equation}
we obtain
\begin{equation}
\hat{H}_\mathrm{FF}(t)=\frac{ds}{dt}\hat{U}_f(t)\hat{H}(s)\hat{U}_f^\dag(t)+i\hbar\bm{(}\partial_t\hat{U}_f(t)\bm{)}\hat{U}_f^\dag(t). 
\end{equation}
In the original proposal~\cite{Takahashi2014}, basis operators were used to express the unitary operator $\hat{U}_f(t)$, but it requires additional conditions to delete influences on measurement outcomes. 
We can avoid requirement of additional conditions by using projection operators, i.e., 
\begin{equation}
\hat{U}_f(t)=e^{i\sum_\sigma f_\sigma(t)\hat{P}_\sigma},\quad\hat{P}_\sigma=|\sigma\rangle\langle\sigma|, 
\label{Eq.uf}
\end{equation}
where $|\sigma\rangle$ is a given measurement basis~\cite{Hatomura2023a}. 
Then, we can obtain the same probability distribution with different timescale
\begin{equation}
|\langle\sigma|\Psi_\mathrm{FF}(t)\rangle|^2=|\langle\sigma|\Psi(s)\rangle|^2. 
\end{equation}
Note again that the measurement basis $|\sigma\rangle$ may depend on time.

%
%
\section{\label{Sec.rel.STA}Relations between different methods}

In this section, we discussion relations between different methods of shortcuts to adiabaticity. 
In particular, we revisit relations between fast-forward scaling and other methods of shortcuts to adiabaticity by using the formulation in Sec.~\ref{Sec.FF.gen}.

%
%
\subsection{Lewis-Riesenfeld theory and counterdiabatic driving}

First, we discuss relation between counterdiabatic driving and the Lewis-Riesenfeld theory, which was discussed in Ref.~\cite{Chen2011,Takahashi2017a}.

%
%
\subsubsection{Dynamical invariant in counterdiabatic driving}

For the total Hamiltonian of counterdiabatic driving (\ref{Eq.adham}), the following Hermitian operator
\begin{equation}
\hat{F}(t)=\sum_n\bar{f}_n|n(t)\rangle\langle n(t)|,
\end{equation}
with an arbitrary constant $\bar{f}_n$ is the dynamical invariant~\cite{Chen2011}. 
It means that the eigenvectors of the dynamical invariant are identical with the eigenstates of the reference Hamiltonian. 
Then, the first term in the Lewis-Riesenfeld phase (\ref{Eq.LRphase}) gives the geometric phase. 
Because the counterdiabatic Hamiltonian (\ref{Eq.cdham}) is off-diagonal in the adiabatic basis, the second term in the Lewis-Riesenfeld phase (\ref{Eq.LRphase}) gives the dynamical phase. 
As a result, the state (\ref{Eq.sol.Seq}) becomes the adiabatic state (\ref{Eq.ad}).

%
%
\subsubsection{\label{Sec.genham.decom}Hamiltonian in the dynamical-invariant basis}

We expand the Hamiltonian $\hat{H}(t)$ by using the dynamical-invariant basis $\{|\phi_n(t)\rangle\}_{n=1,2,\ldots}$. 
By using Eq.~(\ref{Eq.offele}), we obtain
\begin{equation}
\begin{aligned}
\hat{H}(t)=&\sum_n\langle\phi_n(t)|\hat{H}(t)|\phi_n(t)\rangle|\phi_n(t)\rangle\langle\phi_n(t)| \\
&+i\hbar\sum_{\substack{m,n \\ (m\neq n)}}|\phi_n(t)\rangle\langle\phi_n(t)|\partial_t\phi_m(t)\rangle\langle\phi_m(t)|, 
\end{aligned}
\label{Eq.genham.decomposition}
\end{equation}
that is, any Hamiltonian has the same mathematical structure with the total Hamiltonian of counterdiabatic driving (\ref{Eq.adham})~\cite{Takahashi2017a}. 
Indeed, the first term in Eq.~(\ref{Eq.genham.decomposition}) corresponds to the reference Hamiltonian of counterdiabatic driving and the second term in Eq.~(\ref{Eq.genham.decomposition}) corresponds to the counterdiabatic Hamiltonian (\ref{Eq.cdham}). 
It is also obvious that the dynamical-invariant expression of any dynamics (\ref{Eq.sol.Seq}) with the phase (\ref{Eq.LRphase}) is related to the adiabatic state (\ref{Eq.ad}).

%
%
\subsection{Application of fast-forward scaling to the other methods}

For given shortcuts, we may consider further speedup by using fast-forward scaling. 
In this section, we discuss application of fast-forward scaling to counterdiabatic driving and invariant-based inverse engineering. 
It was first discussed in Ref.~\cite{Takahashi2014}, but we reformulate it by adopting the formulation in Sec.~\ref{Sec.FF.gen}.

%
%
\subsubsection{\label{Sec.FFforCD}Application of fast-forward scaling to counterdiabatic driving}

First, we consider application of fast-forward scaling to counterdiabatic driving. 
The fast-forward state is given by
\begin{equation}
|\Psi_\mathrm{FF}(t)\rangle=\hat{U}_f(t)|\Psi_\mathrm{ad}(s)\rangle,
\end{equation}
and we assume a unitary operator (\ref{Eq.uf}) with the projection operator
\begin{equation}
\hat{P}_n(s)=|n(s)\rangle\langle n(s)|, 
\label{Eq.proj.adiabatic}
\end{equation}
which does not change the amplitude of each adiabatic mode. 
Then, the fast-forward Hamiltonian is given by
\begin{equation}
\hat{H}_\mathrm{FF}(t)=\frac{ds}{dt}\hat{H}_\mathrm{ad}(s)-\hbar\sum_n\frac{df_n(t)}{dt}|n(s)\rangle\langle n(s)|. 
\end{equation}
Namely, we can realize fast-forward of counterdiabatic driving by fast-forwarding and amplifying the counterdiabatic Hamiltonian with arbitrary choice of the diagonal part. 
By setting $\hbar\bm{(}df_n(t)/dt\bm{)}=(ds/dt-1)E_n(s)$, we can find the following fast-forward Hamiltonian
\begin{equation}
\hat{H}_\mathrm{FF}(t)=\hat{H}(s)+\frac{ds}{dt}\hat{H}_\mathrm{cd}(s). 
\end{equation}
Here, the reference Hamiltonian is simply fast-forwarded and the counterdiabatic Hamiltonian is fast-forwarded with amplification.

%
%
\subsubsection{Application of fast-forward scaling to invariant-based inverse engineering}

We can also consider application of fast-forward scaling to invariant-based inverse engineering. 
Here, we consider the fast-forward state (\ref{Eq.FFstate}) and we assume a unitary operator (\ref{Eq.uf}) with the projection operator
\begin{equation}
\hat{P}_{\phi_n}(s)=|\phi_n(s)\rangle\langle\phi_n(s)|, 
\end{equation}
which does not change the amplitude of each dynamical mode. 
The fast-forward Hamiltonian is given by
\begin{equation}
\hat{H}_\mathrm{FF}(t)=\frac{ds}{dt}\hat{H}(s)-\hbar\sum_n\frac{df_{\phi_n}(t)}{dt}|\phi_n(s)\rangle\langle\phi_n(s)|. 
\end{equation}
Namely, the off-diagonal part is fast-forwarded and amplified, but choice of the diagonal part is arbitrary.

%
%
\subsection{Fast-forward scaling and counterdiabatic driving}

In this section, we discuss relation between fast-forward scaling and counterdiabatic driving. 
We apply fast-forward scaling to adiabatic time evolution and show its equivalence with counterdiabatic driving. 
It was discussed in Ref.~\cite{Setiawan2017}, but we revisit this problem by adopting the formulation in Sec.~\ref{Sec.FF.gen}. 
We also consider application of fast-forward scaling to nonadiabatic transitions and show its asymptotic equivalence with counterdiabatic driving in the adiabatic limit of reference dynamics~\cite{Hatomura2023a}.

%
%
\subsubsection{\label{Sec.FF.ad}Fast-forward scaling of adiabatic time evolution}

We consider fast-forward scaling of adiabatic time evolution. 
Since the adiabatic state (\ref{Eq.ad}) is not the solution of the Schr\"odinger equation under the reference Hamiltonian, a trick is necessary for application of fast-forward scaling to adiabatic time evolution. 
This trick is known as regularization~\cite{Masuda2010,Setiawan2017}.

The adiabatic state is the solution of the Schr\"odinger equation under the Hamiltonian (\ref{Eq.adham}) and the counterdiabatic Hamiltonian can be expressed as Eq.~(\ref{Eq.cdham.AGP}). 
By considering the adiabatic limit of the reference Hamiltonian, i.e., $d\bm{\lambda}/dt=\bm{\epsilon}$ with a small constant vector $\bm{\epsilon}$, we obtain the regularized Hamiltonian
\begin{equation}
\hat{H}_\mathrm{reg}(t)=\hat{H}(\bm{\lambda})+\bm{\epsilon}\cdot\hat{\mathcal{A}}(\bm{\lambda}), 
\end{equation}
which is asymptotically equal to the reference Hamiltonian and generates the adiabatic state (\ref{Eq.ad}). 
The second term is the regularization term.

Application of fast-forward scaling to this state is equivalent with discussion in Sec.~\ref{Sec.FFforCD}, i.e., we obtain
\begin{equation}
\hat{H}_\mathrm{FF}(t)=\hat{H}\bm{(}\bm{\lambda}(s)\bm{)}+\frac{ds}{dt}\bm{\epsilon}\cdot\hat{\mathcal{A}}\bm{(}\bm{\lambda}(s)\bm{)}. 
\label{Eq.FF.reg}
\end{equation}
The coefficient of the adiabatic gauge potential in Eq.~(\ref{Eq.FF.reg}) becomes finite when the time-rescaling rate $ds/dt$ scales as $\mathcal{O}(\epsilon^{-1})$ because we speedup adiabatic time evolution whose speed is given by $\mathcal{O}(\epsilon)$ with $\epsilon=|\bm{\epsilon}|$. 
This is a simple understanding of the finding in Ref.~\cite{Setiawan2017} and it is nothing but counterdiabatic driving.

%
%
\subsubsection{Fast-forward scaling of nonadiabatic transitions and the adiabatic limit of reference dynamics}

We consider fast-forward scaling of nonadiabatic transitions and the adiabatic limit of reference dynamics~\cite{Hatomura2023a}. 
As mentioned in Sec.~\ref{Sec.nad}, the amount of nonadiabatic transitions is characterized by the coefficients of the adiabatic basis. 
Therefore, we consider the fast-forward state (\ref{Eq.FFstate}) with the projection operator of the adiabatic basis (\ref{Eq.proj.adiabatic}). 
Because we would like to discuss relation between fast-forward scaling and counterdiabatic driving, we set the phase $f_n(t)$ so that the diagonal part of the fast-forward Hamiltonian becomes the fast-forwarded reference Hamiltonian $\hat{H}(s)$. 
Then, we obtain the following fast-forward Hamiltonian
\begin{equation}
\hat{H}_\mathrm{FF}(t)=\hat{H}_\mathrm{ref}(s)+\frac{ds}{dt}\bm{(}\hat{H}_\mathrm{cd}(s)+\hat{H}_\mathrm{nad}(t)\bm{)}, 
\end{equation}
where
\begin{equation}
\begin{aligned}
\hat{H}_\mathrm{nad}(t)=-i\hbar\sum_{\substack{m,n \\ (m\neq n)}}&e^{-i\bm{(}f_m(t)-f_n(t)\bm{)}}\\
&\times|m(s)\rangle\langle m(s)|\partial_sn(s)\rangle\langle n(s)|. 
\end{aligned}
\label{Eq.nadham}
\end{equation}
with $\hbar\bm{(}df_n(t)/dt\bm{)}=(1-ds/dt)E_n(s)$.

Now we discuss the adiabatic limit of reference dynamics. 
For simplicity, we assume uniform fast-forwarding $s=(T_\mathrm{ref}/T_\mathrm{FF})t$, where $T_\mathrm{ref}$ is the operation time of reference dynamics and $T_\mathrm{FF}$ is that of fast-forwarded dynamics. 
Because $T_\mathrm{ref}/T_\mathrm{FF}\gg1$, the dominant term in the phase factor
\begin{equation}
\begin{aligned}
e^{-i\bm{(}f_m(t)-f_n(t)\bm{)}}=&e^{i\int_0^tdt^\prime(T_\mathrm{ref}/T_\mathrm{FF})[E_m\bm{(}s(t^\prime)\bm{)}-E_n\bm{(}s(t^\prime)\bm{)}]}\\
&e^{-i\int_0^tdt^\prime[E_m\bm{(}s(t^\prime)\bm{)}-E_n\bm{(}s(t^\prime)\bm{)}]},
\end{aligned}
\end{equation}
is the first phase factor. 
It gives fast oscillation when the energy gap is large. 
Moreover, it also gives fast oscillation even when the energy gap is small. 
Indeed, $T_\mathrm{ref}$ is larger than the adiabatic timescale $T_\mathrm{ad}$, which is minimal time for the adiabatic condition (\ref{Eq.adcond}) and, roughly speaking, is proportional to the inverse square of the energy gap. 
As a result, the additional term (\ref{Eq.nadham}) effectively vanishes in the time-evolution operator because of fast oscillation. 
Namely, fast-forward scaling of nonadiabatic transitions is asymptotically equivalent with counterdiabatic driving.

%
%
\section{\label{Sec.ver.app}Versatile approximations}

Realization of exact shortcuts to adiabaticity has various difficulties~\cite{Guery-Odelin2019}. 
For example, construction of a counterdiabatic Hamiltonian with the expression~(\ref{Eq.cdham}) requires knowledge of energy eigenstates. 
As mentioned in Sec.~\ref{Sec.intro} and explained in the following subsections, the variational approach and related approaches can (partly) resolve this problem, but implementation is still difficult because counterdiabatic Hamiltonians generally consist of many-body and nonlocal interactions. 
Simaltaneous time-dependent control of a reference Hamiltonian and a counterdiabatic Hamiltonian might also be challenging.

Here we introduce some recent progress in approximate construction and implementation of counterdiabatic driving. 
We also introduce a method for evaluating performance of approximate shortcuts to adiabaticity.

%
%
\subsection{Approximate consrtuction of counterdiabatic driving}

Compared with invariant-based inverse engineering and fast-forward scaling, many general approaches to approximate construction of counterdiabatic driving have been proposed. 
The variational approach~\cite{Sels2017} is one of the significant approaches. 
As mentioned in Sec.~\ref{Sec.intro}, we use a condition for the counterdiabatic Hamiltonian~\cite{Jarzynski2013} as a cost function and optimize a trial counterdiabatic Hamiltonian by minimizing it. 
Recently, direct approaches to the condition for the counterdiabatic Hamiltonian~\cite{Hatomura2021,Takahashi2023,Bhattacharjee2023} and detailed analysis of the variational approach~\cite{Xie2022} were considered. 
These results give a set of linear equations
\begin{equation}
B\vec{a}=\vec{u},
\label{Eq.lineq}
\end{equation}
for approximate counterdiabatic Hamiltonians, where $\vec{a}$ is a coefficient of counterdiabatic terms. 
A matrix $B$ and a vector $\vec{u}$ are given without knowledge of energy eigenstates. 
We will show a unified viewpoint of these approaches below. 
For unified understanding, we use the Frobenius inner product (divided by dimension $D$)
\begin{equation}
(\hat{X}|\hat{Y})=\frac{1}{D}\mathrm{Tr}(\hat{X}^\dag\hat{Y}), 
\end{equation}
between two operators $\hat{X}$ and $\hat{Y}$. 
The Frobenius norm (divided by the square root of dimension) is also defined as
\begin{equation}
\|\hat{X}\|=\sqrt{(\hat{X}|\hat{X})}. 
\end{equation}

%
%
\subsubsection{\label{Sec.variationalCD}Variational approach}

The expression of the counterdiabatic Hamiltonian (\ref{Eq.cdham}) requires the energy eigenstates of the reference Hamiltonian for its construction. 
Here, we discuss the variational approach which enables us to construct approximate counterdiabatic Hamiltonians without knowing energy eigenstates~\cite{Sels2017}. 
This approach is based on the fact that the counterdiabatic Hamiltonian (\ref{Eq.cdham}) satisfies the following condition~\cite{Jarzynski2013,Sels2017}
\begin{equation}
\mathcal{L}_H\bm{(}i\hbar\partial_t\hat{H}(t)+\mathcal{L}_H\hat{H}_\mathrm{cd}(t)\bm{)}=0. 
\label{Eq.cond.cdham}
\end{equation}
An Hermitian operator which satisfies this condition instead of the counterdiabatic Hamiltonian is equal to the counterdiabatic Hamiltonian (\ref{Eq.cdham}) except for the diagonal elements, i.e., such an operator cancels out nonadiabatic transitions although phase factors of a resulting state may be different from the adiabatic state (\ref{Eq.ad}).

In the variational approach, we introduce the following operator
\begin{equation}
\hat{G}[\hat{H}_\mathrm{cd}^\mathrm{tri}(t)]=\partial_t\hat{H}(t)-\frac{i}{\hbar}\mathcal{L}_H\hat{H}_\mathrm{cd}^\mathrm{tri}(t),
\label{Eq.Gfunc}
\end{equation}
where $\hat{H}_\mathrm{cd}^\mathrm{tri}(t)$ is a trial counterdiabatic Hamiltonian. 
Then, we minimize the action
\begin{equation}
S[\hat{H}_\mathrm{cd}^\mathrm{tri}(t)]=\|G^2[\hat{H}_\mathrm{cd}^\mathrm{tri}(t)]\|^2,
\end{equation}
against the trial counterdiabatic Hamiltonian by calculating the variational principle
\begin{equation}
\frac{\delta S[\hat{H}_\mathrm{cd}^\mathrm{tri}(t)]}{\delta\hat{H}_\mathrm{cd}^\mathrm{tri}(t)}=0. 
\label{Eq.vari.op}
\end{equation}
As a result, we find time dependence of the trial counterdiabatic Hamiltonian. 
When the trial counterdiabatic Hamiltonian consists of operators which are identical with those in the exact counterdiabatic Hamiltonian (\ref{Eq.cdham}), this method can reproduce the exact counterdiabatic Hamiltonian. 
Ohterwise, it gives an approximate counterdiabatic Hamiltonian, which may be useful for suppressing some nonadiabatic transitions.

An important task of the variational approach is determination of operators for the trial counterdiabatic Hamiltonian. 
An integral expression of the counterdiabatic Hamiltonian~\cite{Claeys2019} gives an insight into this task. 
We can rewrite the counterdiabatic Hamiltonian (\ref{Eq.cdham}) as
\begin{equation}
\begin{aligned}
\hat{H}_\mathrm{cd}(t)=-\frac{1}{2}\lim_{\eta\to0}\int_{-\infty}^\infty du\ &\mathrm{sgn}(u)e^{-\eta|u|}\\
&\times\hat{U}_\mathrm{fic}^\dag(u)\bm{(}\partial_t\hat{H}(t)\bm{)}\hat{U}_\mathrm{fic}(u),
\end{aligned}
\label{Eq.cdham.int}
\end{equation}
where $\hat{U}_\mathrm{fic}(u)$ is a fictitious time evolution operator $\hat{U}_\mathrm{fic}(u)=e^{-\frac{i}{\hbar}\hat{H}(t)u}$~\cite{Claeys2019}. 
We can easily confirm correctness of this expression by calculating the matrix element $\langle m(t)|\hat{H}_\mathrm{cd}(t)|n(t)\rangle$. 
By using the Baker-Campbell-Hausdorff formula
\begin{equation}
\mathrm{Ad}_{e^{\hat{X}}}\hat{Y}=e^{\mathrm{ad}_{\hat{X}}}\hat{Y}=\sum_{k=0}^\infty\frac{1}{k!}\mathrm{ad}_{\hat{X}}^k\hat{Y},
\label{Eq.BCHformula}
\end{equation}
where $\mathrm{Ad}_{e^{\hat{X}}}\hat{Y}=e^{\hat{X}}\hat{Y}e^{-\hat{X}}$ is the adjoint action of a Lie group on its Lie algebra and $\mathrm{ad}_{\hat{X}}\hat{Y}=[\hat{X},\hat{Y}]$ is the adjoint action of the Lie algebra, we can expand the integrand of Eq.~(\ref{Eq.cdham.int}) in terms of the nested commutators 
\begin{equation}
\hat{O}_k=\mathcal{L}_H^k\partial_t\hat{H}(t),\quad k=0,1,2,\ldots. 
\label{Eq.nested.basis}
\end{equation}
Rigorously speaking, the series in Eq.~(\ref{Eq.BCHformula}) do not converge in general, and thus we can not exchange summation and integral. 
However, once we assume it, we notice that the counterdiabatic Hamiltonian consists of the odd nested commutators, $\hat{O}_{2k-1}$ $(k=1,2,\ldots)$, because the terms consist of the even nested commutators are odd functions in the integrand. 
Therefore, we can make the following ansatz
\begin{equation}
\hat{H}_\mathrm{cd}^\mathrm{tri}(t)=i\hbar\sum_{k=1}^{K_\mathrm{tr}}a_k^\mathrm{tri}(t)\hat{O}_{2k-1}, 
\label{Eq.nested.ansatz}
\end{equation}
where $\{a_k^\mathrm{tri}(t)\}_{k=1,2,\ldots,K_\mathrm{tr}}$ is time-dependent coefficients which are scheduled by using the variational principle (\ref{Eq.vari.op}), and $K_\mathrm{tr}$ is an integer for truncation~\cite{Claeys2019}. 
It can be the exact counterdiabatic Hamiltonian, but construction of the exact counterdiabatic Hamiltonian generally requires exponentially large $K_\mathrm{tr}$.

In Ref.~\cite{Xie2022}, the variational operation (\ref{Eq.vari.op}) was analyzed with the ansatz (\ref{Eq.nested.ansatz}) and a set of linear equations (\ref{Eq.lineq}) with
\begin{equation}
B_{kl}=\|\hat{O}_{k+l}\|^2,\quad u_k=-\|\hat{O}_k\|^2,
\label{Eq.bkluk.vari}
\end{equation}
for $a_k=a_k^\mathrm{tri}(t)$ was derived.

%
%
\subsubsection{\label{Sec.algebraicCD}Algebraic approach}

The algebraic approach~\cite{Hatomura2021} directly solves Eq.~(\ref{Eq.cond.cdham}) by using basis operators. 
For a $D$-dimensional quantum system, any operator can be expanded by using some of $(D^2-1)$ basis operators $\{\hat{L}_\mu\}_{\mu=1,2,\ldots,D^2-1}$ and the identity operator $\hat{L}_0=\hat{1}$. 
We assume that the basis operators $\{\hat{L}_\mu\}_{\mu=1,2,\ldots,D^2-1}$ are given by traceless Hermitian operators. 
The basis operators satisfy
\begin{equation}
(\hat{L}_\mu|\hat{L}_\nu)=\delta_{\mu\nu},
\end{equation}
and
\begin{equation}
[\hat{L}_\mu,\hat{L}_\nu]=i\sum_{\lambda}f_{\mu\nu\lambda}\hat{L}_\lambda,
\end{equation}
with an antisymmetric tensor $f_{\mu\nu\lambda}$.

We express the reference Hamiltonian as
\begin{equation}
\hat{H}(t)=\sum_{i=1}^Mh_{i}(t)\hat{L}_{\mu_i},\quad\mu_i\in\{1,2,\ldots,D^2-1\},
\label{Eq.ham.alg}
\end{equation}
with time-dependent coefficients $\{h_{i}(t)\}_{i=1,2,\ldots,M}$, where $M$ is an integer satisfying $M\le D^2-1$. 
Here and hereafter, we omit the identity operator $\hat{L}_0=\hat{1}$ since it only contributes to a global phase factor. 
By calculating the odd nested commutators, $\mathcal{L}_H^{2k-1}\partial_t\hat{H}(t)$ ($k=1,2,\ldots$), with the expression (\ref{Eq.ham.alg}), we can specify basis operators for a trial counterdiabatic Hamiltonian, $\{\hat{L}_{\tilde{\mu}_i}\}_{i=1,2,\ldots,\tilde{M}}$, $\tilde{\mu}_i\in\{1,2,\ldots,D^2-1\}$, with an integer $\tilde{M}$ satisfying $\tilde{M}\le D^2-1$. 
Then, we can express a trial counterdiabatic Hamiltonian as
\begin{equation}
\hat{H}_\mathrm{cd}^\mathrm{tri}(t)=i\hbar\sum_{i=1}^{\tilde{M}}\tilde{a}_{i}^\mathrm{tri}(t)\hat{L}_{\tilde{\mu}_i},
\label{Eq.ansatz.alg}
\end{equation}
with certain time-dependent coefficients $\{\tilde{a}_{i}^\mathrm{tri}(t)\}_{i=1,2,\ldots,\tilde{M}}$.

In the algebraic approach~\cite{Hatomura2021}, we consider the inner product of the condition (\ref{Eq.cond.cdham}) and the basis operators $\{\hat{L}_{\tilde{\mu}_i}\}_{i=1,2,\ldots,\tilde{M}}$. 
As a result, we obtain a set of linear equations (\ref{Eq.lineq}) with
\begin{equation}
B_{kl}=(\mathcal{L}_H\hat{L}_{\tilde{\mu}_k}|\mathcal{L}_H\hat{L}_{\tilde{\mu}_l}),\quad u_k=(\mathcal{L}_H\partial_t\hat{H}(t)|\hat{L}_{\tilde{\mu}_k}),
\label{Eq.lineq.alg}
\end{equation}
for $a_k=\tilde{a}_{k}^\mathrm{tri}(t)$. 
By solving Eq.~(\ref{Eq.lineq}) with Eq.~(\ref{Eq.lineq.alg}), we obtain an approximate counterdiabatic Hamiltonian. 
Note that the algebraic approach is equivalent with the variational approach when we adopt the same basis operators~\cite{Hatomura2021}, i.e., when we make a same ansatz (\ref{Eq.ansatz.alg}), but we can directly obtain the set of linear equations (\ref{Eq.lineq}) which is the result of the variational operation (\ref{Eq.vari.op}). 
The algebraic approach can also give the exact counterdiabatic Hamiltonian when the set of the basis operators in the trial counterdiabatic Hamiltonian is identical with that in the exact counterdiabatic Hamiltonian.

Recently, application of the Lanczos algorithm to construction of the counterdiabatic Hamiltonian was proposed~\cite{Takahashi2023,Bhattacharjee2023}. 
In this approach, we solve Eq.~(\ref{Eq.cond.cdham}) by using the Krylov basis instead of the basis operators. 
The key idea is that the nested commutators naturally generate the Krylov subspace, $\{\mathcal{L}_H^{k}\partial_t\hat{H}(t)\}_{k=0,1,2,\ldots}$, which spans the counterdiabatic Hamiltonian. 
The Krylov basis $\{\hat{\mathcal{O}}_k\}_{k=0,1,2,\ldots,K-1}$ is generated as follows
\begin{equation}
\begin{aligned}
&b_0\hat{\mathcal{O}}_0=\partial_t\hat{H}(t), \\
&b_1\hat{\mathcal{O}}_1=\mathcal{L}_H\hat{\mathcal{O}}_0, \\
&b_k\hat{\mathcal{O}}_k=\mathcal{L}_H\hat{\mathcal{O}}_{k-1}-b_{k-1}\hat{\mathcal{O}}_{k-2},\quad k=2,3,\ldots,K-1,
\end{aligned}
\label{Eq.Kbasis}
\end{equation}
where the coefficient $b_k$ ($k=0,1,2,\ldots,K-1$) is the normalization factor for $\hat{\mathcal{O}}_k$, i.e., it is defined so that $(\hat{\mathcal{O}}_k|\hat{\mathcal{O}}_k)=1$. 
Then, the Krylov basis $\{\hat{\mathcal{O}}_k\}_{k=0,1,2,\ldots,K-1}$ is orthonormal. 
Note that generation of the Krylov basis terminates automatically with $b_K=0$. 
The integer $K$ is known as the Krylov dimension and it satisfies $K\le D^2-D+1$. 
By using the Krylov basis, fictitious time evolution of the operator $\hat{\mathcal{O}}_0=\partial_t\hat{H}(t)/b_0$ in the Heisenberg picture, $\hat{\mathcal{O}}(s)=\hat{U}_\mathrm{fic}^\dag(s)\hat{\mathcal{O}}_0\hat{U}_\mathrm{fic}(s)$, which appears in Eq.~(\ref{Eq.cdham.int}), can be expanded as
\begin{equation}
\hat{\mathcal{O}}(s)=\sum_{k=0}^{K-1}i^k\psi_k(s)\hat{\mathcal{O}}_k,
\end{equation}
with a coefficient $\psi_k(s)$, whose time dependence can be obtained by considering the Heisenberg equation. 
Because $\psi_k(-s)=(-1)^k\psi_k(s)$ holds and we can commute the summation and the integral in Eq.~(\ref{Eq.cdham.int}), the even terms vanish by integration. 
As a result, it can be expressed as
\begin{equation}
\hat{H}_\mathrm{cd}(t)=i\hbar\sum_{k=1}^{\lfloor K/2\rfloor}\bar{a}_k(t)\hat{\mathcal{O}}_{2k-1}, 
\label{Eq.cdham.Krylov}
\end{equation}
where
\begin{equation}
\bar{a}_k(t)=\lim_{\eta\to0}\frac{1}{\hbar}\int_0^\infty dse^{-\eta|s|}b_0(-1)^k\psi_{2k-1}(s). 
\label{Eq.coeff.Krylov}
\end{equation}

Notably, we can determine the coefficients without calculating the above expression (\ref{Eq.coeff.Krylov}). 
As in the case of the algebraic approach~\cite{Hatomura2021}, we consider to directly solve Eq.~(\ref{Eq.cond.cdham}). 
By substituting (\ref{Eq.cdham.Krylov}) for Eq.~(\ref{Eq.cond.cdham}), we obtain a set of linear equations (\ref{Eq.lineq}) with
\begin{equation}
\begin{aligned}
B_{kl}=&(b_{2k-1}^2+b_{2k}^2)\delta_{kl}\\
&+b_{2k-2}b_{2k-1}\delta_{k,l+1}+b_{2k}b_{2k+1}\delta_{k+1,l},
\end{aligned}
\label{Eq.bkl.Krylov}
\end{equation}
and
\begin{equation}
u_k=-b_0b_1\delta_{k1},
\label{Eq.uk.Krylov}
\end{equation}
for $a_k=\bar{a}_{k}(t)$. 
Remarkably, we can efficiently solve the set of the linear equations (\ref{Eq.lineq}) since $B$ is the tri-diagonal matrix, and then we obtain the counterdiabatic Hamiltonian. 
The Krylov dimension $K$ is exponentially large in general. 
Therefore, we may truncate generation of the Krylov basis (\ref{Eq.Kbasis}) and construction of the linear equation (\ref{Eq.lineq}), and then we obtain approximate counterdiabatic Hamiltonians.

%
%
\subsubsection{Unified viewpoint}

Here, we give a unified viewpoint of the above three approaches. 
Equation (\ref{Eq.bkluk.vari}) for the set of the linear equations (\ref{Eq.lineq}) in the variational approach with the nested-commutator ansatz (\ref{Eq.nested.ansatz}) can be rewritten as
\begin{equation}
B_{kl}=(\mathcal{L}_H\hat{O}_{2k-1}|\mathcal{L}_H\hat{O}_{2l-1}),\quad u_k=-(\mathcal{L}_H\partial_t\hat{H}(t)|\hat{O}_{2k-1}), 
\end{equation}
and Eqs.~(\ref{Eq.bkl.Krylov}) and (\ref{Eq.uk.Krylov}) for it in the Krylov approach can be rewritten as
\begin{equation}
B_{kl}=(\mathcal{L}_H\hat{\mathcal{O}}_{2k-1}|\mathcal{L}_H\hat{\mathcal{O}}_{2l-1}),\quad u_k=-(\mathcal{L}_H\partial_t\hat{H}(t)|\hat{\mathcal{O}}_{2k-1}). 
\end{equation}
Note that $\hat{O}_k$ is the nested commutator (\ref{Eq.nested.basis}) and $\hat{\mathcal{O}}_k$ is the Krylov basis (\ref{Eq.Kbasis}). 
We also find a similar expression for the algebraic approach in Eq.~(\ref{Eq.lineq.alg}). 
Namely, these three approaches have the same mathematical structure. 
Note that the basis in the algebraic approach is Hermitian, whereas these in the other approaches are anti-Hermitian, and thus the sign in $u_k$ is different.

The variational approach with the nested-commutator ansatz~\cite{Xie2022} may be advantageous because Eq.~(\ref{Eq.bkluk.vari}) can simply be calculated in order, but the matrix $B$ may be dense. 
The algebraic approach~\cite{Hatomura2021} may be advantageous because we can intuitively specify basis operators in some cases, e.g., multiple Pauli matrices with odd number of the Pauli-Y in multi-qubits systems consisting of the Pauli-X and the Pauli-Z. 
The Krylov approach~\cite{Takahashi2023,Bhattacharjee2023} is advantageous because the matrix $B$ is the tri-diagonal matrix and the set of the linear equations (\ref{Eq.lineq}) can effectively be soloved, but the generation of the Krylov basis and determination of the coefficients may be complicated in some cases. 
We can use appropriate one depending on given situations.

%
%
\subsubsection{\label{Sec.appCD.appli}Applications}

The above mentioned approaches open the possibility of counterdiabatic driving applied to many-body systems and quenched disorder systems. 
Namely, these approaches have potential to improve performance of quantum annealing, quantum heat engines, many-body state preparation, etc. 
Indeed, various applications of the variational approach have been reported. 
It is also notable that the variational approach was also extended to classical systems~\cite{Gjonbalaj2022}.

The variational approach was applied to quantum annealing in the Lechner-Hauke-Zoller architecture~\cite{Hartmann2019a}, the $p$-spin model~\cite{Passarelli2020}, the $p$-spin model with two-parameter counterdiabatic terms~\cite{Prielinger2021}, and spin models with local and two-body counterdiabatic terms~\cite{Hartmann2022}. 
It was also applied to reverse annealing~\cite{Passarelli2023}, which is a variant of quantum annealing. 
Rotated ansatz, in which unwanted driving terms are canceled out by moving to a rotating frame, was proposed for experimantal realization~\cite{Mbeng2022}. 
Speedup is not in an exponential way, but fidelity to the ground state was enhanced in all the examples.

Application of the variational approach to quantum heat engines was also discussed. 
As working mediums, Ising spin models were used~\cite{Hartmann2020,Hartmann2020a}. 
Performance improvement by higher-order counterdiabatic terms was also confirmed~\cite{Hartmann2020a}.

State preparation is an important process for quantum algorithms. 
Preparation of long-range topological order in the honeycomb Kitaev model was considered~\cite{Kumar2021} and state transfer on a Heisenberg spin chain was discussed~\cite{Ji2022}.

%
%
\subsection{\label{Sec.digCD}Digital implementation of counterdiabatic driving}

In the previous section, we discussed approximate construction of counterdiabatic driving. 
Here, we consider approximate implementation of counterdiabatic driving.

Realization of counterdiabatic driving requires simultaneous control of different time-dependent interactions, i.e., the reference Hamiltonian and the counterdiabatic Hamiltonian. 
This difficulty motivates us to apply digital quantum simulation (Hamiltonian simulation) to counterdiabatic driving. 
In digital quantum simulation, we devide time-evolution operators into sequences of quantum gate operations~\cite{Huyghebaert1990,Lloyd1996}. 
Namely, we can separately control the reference Hamiltonian and the counterdiabatic Hamiltonian in digitized counterdiabatic driving. 
Moreover, a problem of time-dependent control can be translated into that of a time duration with constant interaction.

The simplest way of digitized counterdiabatic driving is application of the Lie-Trotter formula~\cite{Trotter1959} to the time-evolution operator of counterdiabatic driving. 
Namely, we approximate it as
\begin{equation}
\hat{U}_\mathrm{ad}(T)\approx\prod_{n=M}^1\left(e^{-\frac{i}{\hbar}\frac{T}{M}\hat{H}(nT/M)}e^{-\frac{i}{\hbar}\frac{T}{M}\hat{H}_\mathrm{cd}(nT/M)}\right),
\end{equation}
where $T$ is the final time and $M$ is the number of time slices. 
Further decomposition of the reference Hamiltonian and the counterdiabatic Hamiltonian may also be considered for experimental realization.

Digitized counterdiabatic driving was implemented on superconducting and ion-trap quantum processors~\cite{Hegade2021,Hegade2022}. 
Here, state preparation of entangled states was demonstrated~\cite{Hegade2021} and quantum annealing was considered~\cite{Hegade2022}. 
Portfolio optimization problem~\cite{Hegade2022a} and factorization~\cite{Hegade2023} was discussed as practical applications. 
Moreover, unusual error scaling was pointed out~\cite{Hatomura2023} and this theory was extended to digitization of general dynamics~\cite{Hatomura2023b}.

Note that we can also consider other decomposition. 
For example, use of a product formula for commutators enables us to precisely and efficiently realize approximate counterdiabatic Hamiltonians~\cite{Chen2022}.

%
%
\subsection{\label{Sec.performance}Performance evaluation by quantum speed limits}

Owing to difficulties in exact construction and implementation of shortcuts to adiabaticity, some approximations are usually required. 
It may be possible to directly compare approximate results with ideal ones in some cases, but in general accessible information is limited. 
It is an important task to evaluate performance of approximate shortcuts to adiabaticity with partial information.

For given two quantum dynamics $|\Psi_1(t)\rangle$ and $|\Psi_2(t)\rangle$ which are governed by Hamiltonians $\hat{H}_1(t)$ and $\hat{H}_2(t)$, an inequality
\begin{equation}
|\langle\Psi_1(t)|\Psi_2(t)\rangle|\ge\cos\left(\frac{1}{\hbar}\int_0^tdt^\prime\mathcal{L}(t^\prime)\right),
\label{Eq.ineq.ST}
\end{equation}
where
\begin{equation}
\mathcal{L}(t)=\sigma[\hat{H}_1(t^\prime)-\hat{H}_2(t^\prime),|\Psi_i(t^\prime)\rangle],
\label{Eq.integrand.ST}
\end{equation}
holds~\cite{Suzuki2020}. 
Here, the index $i$ can be either $1$ or $2$, and $\sigma[\hat{X},|\Psi\rangle]$ is the standard deviation, $\sigma[\hat{X},|\Psi\rangle]=\sqrt{\langle\Psi|\hat{X}^2|\Psi\rangle-\langle\Psi|\hat{X}|\Psi\rangle^2}$. 
The left-hand side of this inequality is the overlap between two dynamics, and the right-hand side of it represent how difference between two Hamiltonians affect dynamics.

We can also derive a discretized version of this inequality~\cite{Hatomura2022}
\begin{equation}
|\langle\Psi_1(T)|\Psi_2(T)\rangle|\ge\cos\left(\sum_{n=1}^M\mathcal{L}_n\right),
\label{Eq.ineq.H}
\end{equation}
where
\begin{equation}
\begin{aligned}
\mathcal{L}_n=\arccos|&\langle\Psi_i(nT/M)|\hat{U}_{\bar{i}}\bm{(}nT/M,(n-1)T/M\bm{)}\\
&\times[\hat{U}_i\bm{(}nT/M,(n-1)T/M\bm{)}]^\dag|\Psi_i(nT/M)\rangle|. 
\end{aligned}
\end{equation}
Here, $\hat{U}_i\bm{(}nT/M,(n-1)T/M\bm{)}$ is a discritized time evolution operator for dynamics $|\Psi_i(t)\rangle$ from time $(n-1)T/M$ to $nT/M$, and $\bar{i}$ is $1$ or $2$ for $i=2$ or $i=1$, respectively. 
In this case, the right-hand side of the inequality represents how difference between two time-evolution operators affect dynamics.

These inequalities can be used to evaluate performance of quantum control. 
Indeed, we can evaluate the worst-case overlap between ideal dynamics and real dynamics by using information of Hamiltonians (time-evolution operators) and one of two dynamics.

In Ref.~\cite{Suzuki2020}, the inequality (\ref{Eq.ineq.ST}) was derived to evaluate performance of finite-time adiabatic control. 
Here, one of dynamics is the adiabatic state (\ref{Eq.ad}), and the other dynamics is the solution of the Schr\"odinger equation (\ref{Eq.Seq}). 
Then, we find that the overlap between the adiabatic state and real dynamics can be evaluated by using the standard deviation of the counterdiabatic Hamiltonian with the adiabatic state. 
Remarkably, when we consider adiabatic control with a single energy eigenstate $|n(\bm{\lambda})\rangle$, it is given by
\begin{equation}
\mathcal{L}(t)=\hbar\sqrt{\sum_{i,j}g_{ij}(\bm{\lambda})\dot{\lambda}_i\dot{\lambda}_j},
\label{Eq.Lt.QGT}
\end{equation}
where
\begin{equation}
g_{ij}(\bm{\lambda})=\langle\partial_in(\bm{\lambda})|\bm{(}1-|n(\bm{\lambda})\rangle\langle n(\bm{\lambda})|\bm{)}|\partial_jn(\bm{\lambda})\rangle,
\end{equation}
with $\partial_i=\partial/\partial\lambda_i$ is known as the quantum geometric tensor of the $n$th eigenstate. 
Remarkably, time integral of the quantity (\ref{Eq.Lt.QGT}) does not depend on time, but it depends on a path in parameter space. 
It is generally large, and thus the inequality itself (\ref{Eq.ineq.ST}) may not be useful~\cite{Suzuki2020,Hatomura2020}. 
Even such a situation, the quantity (\ref{Eq.integrand.ST}) can be used as a measure of potential nonadiabaticity~\cite{Suzuki2020}.

In Ref.~\cite{Hatomura2021}, the inequality~(\ref{Eq.ineq.ST}) was used to evaluate performance of approximate counterdiabatic driving. 
In this case, one of dynamics is the adiabatic state (\ref{Eq.ad}) and the other dynamics is a state driven by approximate counterdiabatic driving. 
The overlap between two dynamics can be evaluated by using the standard deviation of difference between the exact counterdiabatic Hamiltonian and an approximate counterdiabatic Hamiltonian with the adiabatic state. 
When an approximate counterdiabatic Hamiltonian is given by a product of $\dot{\bm{\lambda}}$ and a parameter-dependent operator, i.e., it implicitly depends on time through the parameter $\bm{\lambda}$, time integral of the quantity (\ref{Eq.integrand.ST}) also becomes path-dependent quantity. 
Compared with performance evaluation of adiabatic control, the inequality can be tight even for many-body systems~\cite{Hatomura2021}. 
Notably, we can also evaluate influence of a bath by regarding it as approximation~\cite{Funo2021}.

In Ref.~\cite{Hatomura2022b}, the inequality~(\ref{Eq.ineq.ST}) was used to evaluate performance of approximate invariant-based inverse engineering. 
Since it is a hard task to find dynamical invariants of given systems, we usually consider approximate Hamiltonians, for which we can find dynamical invariants. 
We can design target dynamics with dynamical invariants of approximate Hamiltonians and find parameter schedules by using invariant-based inverse engineering. 
Then, we can apply these parameter schedules to real systems. 
In this case one of dynamics is target dynamics and the other dynamics is real dynamics. 
The overlap between two dynamics can be evaluated by using the standard deviation of difference between the real Hamiltonian and the approximate Hamiltonian with the designed target dynamics. 
Note that we can also apply the same idea to fast-forward scaling. 
Indeed, we also knows target dynamics in fast-forward scaling.

In Ref.~\cite{Hatomura2023}, the inequality~(\ref{Eq.ineq.H}) was used to evaluate performance of digitized counterdiabatic driving. 
In this case, one of dynamics is the adiabatic state (\ref{Eq.ad}) and the other dynamics is a digitized dynamics assisted by the counterdiabatic Hamiltonian. 
Remarkably, scaling of errors $\mathcal{O}(T/M^2)$ was found, whereas the conventional theory predicts $\mathcal{O}(T^2/M)$~\cite{Huyghebaert1990,Lloyd1996}. 
Moreover, we can find the same error scaling in decomposition of general dynamics by considering the dynamical-invariant basis~\cite{Hatomura2023b}.

%
%
\section{\label{Sec.progress}Other recent progress}

We explained recent progress on versatile approximate approaches to counterdiabatic driving above. 
In this section, we introduce other recent progress on shortcuts to adiabaticity after publication of the systematic review in 2019~\cite{Guery-Odelin2019}.

%
%
\subsection{Performance improvement}

The quantum approximate optimization algorithm~\cite{Farhi2014,Zhou2020}, also known as the quantum alternating operator ansatz~\cite{Hadfield2019}, (QAOA) is another notable application of NISQ devices. 
In QAOA, we seek for the ground state of a given problem Hamiltonian by applying the problem Hamiltonian and mixer Hamiltonians in alternating ways and by optimizing their coefficients by using classical computers. 
It can naturally be associated with digital quantum simulation of quantum annealing~\cite{Kadowaki1998}, or more generally adiabatic quantum computation~\cite{Farhi2000}.

Utilization of counterdiabatic driving for QAOA was proposed~\cite{Yao2021,Wurtz2022,Chandarana2022}. 
The idea of counterdiabatic QAOA (CD-QAOA) is two-fold. 
On the one hand, counterdiabatic driving has ability to shorten operation time, and thus it naturally results in reduction of unitary operators for QAOA in accordance with the analogy to digital quantum simulation of adiabatic quantum computation. 
On the other hand, the counterdiabatic Hamiltonians cancel out nonadiabatic transitions, which are transitions between different energy levels and take place when energy eigenstates quickly change with energy gap closing, and thus they have strong ability to introduce changes for given states. 
Mechanisms of CD-QAOA were studied in Ref.~\cite{Wurtz2022}. 
Performance of CD-QAOA depends on choice of initial parameters and classical optimizers. 
Reinforcement learning was used for optimization~\cite{Yao2021}, and meta-learning technique using recurrent neural networks was considered for initial parameter setting~\cite{Chandarana2022a}. 
As other development, power of single-layer QAOA~\cite{Guan2023} and improvement by higher-order counterdiabatic terms~\cite{Vizzuso2024} were confirmed, and the theory was extended to photonic systems with continuous variables~\cite{Chandarana2023a}. 
As practical applications, portfolio optimization~\cite{Hegade2022a} and protein folding~\cite{Chandarana2023} were discussed.

It is also an important task to improve performance of approximate shortcuts to adiabaticity. 
From the early stage, numerical optimization of local approximate counterdiabatic Hamiltonians has been discussed~\cite{Saberi2014,Campbell2015}. 
In this direction, various approaches have been reported. 
Gradient ascent pulse engineering (GRAPE) is a method of pulse shaping to obtain a target state, where pulses are optimized by applying gradient ascent to the overlap between the target state and the final state. 
Enhanced shortcuts to adiabaticity (eSTA) was proposed as a method to enhance performance of given approximate shortcuts by using GRAPE~\cite{Whitty2020}. 
Robustness of eSTA applied to atom transport was confirmed~\cite{Whitty2022} and higher-order generalization was proposed~\cite{Whitty2022a}. 
It was applied to state preparation of a spin squeezed state~\cite{Odelli2023}. 
In optimal control theory, we introduce additional driving and consider optimization of it. 
Counterdiabatic optimized local driving (COLD) was proposed as a combination of optimal control theory with local approximate counterdiabatic driving~\cite{Cepaite2023}. 
Benchmark test was performed by using nontrivial spin systems~\cite{Barone2023}. 
The Pauli-Y matrix is known as the first-order counterdibatic term in local approximate counterdiabatic driving of quantum annealing. 
Greedy parameter optimization was considered to determine the sign of the Pauli-Y term~\cite{Kadowaki2023}. 
Tensor network is a powful tool of classical simulation for quantum many-body systems. 
It was applied to construction of approximate counterdiabatic Hamiltonians~\cite{Kim2023,Keever2023}.

Performance is not determined by only considering fidelity and speed, but cost is also an important measure of performance. 
Speedup via shortcuts to adiabaticity is not free. 
Indeed, we have to pay enough energy cost for it. 
Tradeoff between speed and cost of shortcuts to adiabaticity has been debated for a long time~\cite{Santos2015,Zheng2016,Campbell2017,Funo2017,Torrontegui2017,Tobalina2018,Abah2019}. 
Recently application of counterdiabatic driving only for the impulse regime was proposed as an energy-saving approach~\cite{Carolan2022}. 
It was shown that application of counterdiabatic driving in the vicinity of energy gap closing is enough for suppressing nonadiabatic transitions and reduction of driving saves energy costs. 
An idea of energy costs was also applied to fast-forward scaling. 
By tailoring phase degrees of freedom in fast-forward scaling, we can realize energy-saving and time-saving control~\cite{Hatomura2024}.

%
%
\subsection{Exploring quantum systems}

Use of the counterdiabatic Hamiltonian is not limited to suppression of nonadiabatic transitions. 
As naturally expected, information of eigen-energies and eigenstates is encoded on the counterdiabatic Hamiltonian. 
Indeed, the counterdiabatic Hamiltonian is related to the quantum geometric tensor~\cite{delCampo2012,Funo2017}. 
Therefore, it can be used as a probe for exploring related phenomena. 
Remarkably, even approximate counterdiabatic Hamiltonians work well as probes.

Nonadiabatic transitions and resulting defects reflect critical exponents of driven systems, which is known as the Kibble-Zurek scaling. 
The Kibble-Zurek scaling was explored through counterdiabatic Hamiltonians~\cite{Puebla2020}. 
Transitions from integrable to chaotic change properties of energy spectra. 
It was shown that counterdiabatic Hamiltonians can be used as sensitive probes for detecting these transitions~\cite{Pandey2020,Bhattacharjee2023}. 
It is also possible to study landscape of energy eigenstates in paramter space. 
Such adiabatic landscape gives optimal control schemes in parameter space~\cite{Sugiura2021}. 
In many-body systems, energy gap closing is related with quantum phase transitions. 
It was shown that approximate counterdiabatic Hamiltonians can be used as probes for quantum phase transitions~\cite{Hatomura2021,Kim2023}.

%
%
\subsection{\label{Sec.open}Open quantum systems}

Application of shortcuts to adiabaticity is not limited to closed quantum systems. 
Indeed, theory of counterdiabatic driving was extended to open quantum systems.

The Lindblad equation is one of the fundamental equations of motion in open quantum systems. 
By using basis operators, we can give matrix expression of the Lindblad superoperator, and the Lindblad equation is given in the form of the Schr\"odinger equation with a non-Hermitian Hamiltonian. 
Then, the counterdiabatic Hamiltonian was derived by analogy to that of closed quantum systems~\cite{Vacanti2014,Santos2021}. 
Similarly, the variational approach was extended to open quantum systems~\cite{Passarelli2022}.

As mentioned in Sec.~\ref{Sec.LR}, the density operator is a tirivial example of the dynamical invariant in closed quantum mechanics, and thus its eigenvalues are independent of time. 
In contrast, they depend on time in open quantum systems. 
It was shown that time-dependent eigenvalues give a non-Hermitian part, and remaining Hermitian part is given in the form of Eq.~(\ref{Eq.genham.decomposition})~\cite{Alipour2020}.

An idea of inverse engineering is also available in open quantum systems. 
By substituting given target dynamics for master equations, parameter schedules for realizing target dynamics can be obtained~\cite{Dann2019,Alipour2020,Dupays2020}. 
In Ref.~\cite{Dann2019,Dupays2020}, inverse engineering of fast equilibration (thermalization) was discussed. 
Speedup of cooling, heating, and isothermal strokes was demonstrated in Ref.~\cite{Alipour2020}. 
It was applied to performance enhancement of quantum heat engines~\cite{Pedram2023}.

Composite systems can be regarded as open quantum systems. 
By engineering an ancilla system, counterdiabatic driving can be realized in a target system~\cite{Touil2021}. 
Continuously monitored systems can also be viewed as open quantum systems. 
Enhancement of the quantum Zeno dragging was proposed by utilizing the idea of counterdiabatic driving~\cite{Lewalle2023}.

%
%
\section{Conclusion remarks}

We gave the overview on the background and theory of shortcuts to adiabaticity. 
Relations between different methods of shortcuts to adiabaticity were revisited. 
In particular, relations between fast-forward scaling and other methods of shortcuts to adiabaticity were reformulated. 
Various versatile approximations for shortcuts to adiabaticity were introduced. 
Particularly, the unified viewpoint of versatile approaches to counterdiabatic driving was discussed. 
Some recent progress was also summarized.

In this Topical Review, we did not consider concrete examples because we would like to clearly emphasize general and universal properties of shortcuts to adiabaticity. 
For readers who are interested in simple examples, a topical review~\cite{Masuda2022} and pedagogical introductions~\cite{Nakahara2022,Takahashi2022} are useful. 
Regarding approximate shortcuts to adiabaticity for complicated examples, comprehensive systematic reviews~\cite{Torrontegui2013,Guery-Odelin2019} are available.

\begin{acknowledgments}
This work was supported by JST Moonshot R\&D Grant Number JPMJMS2061. 
\end{acknowledgments}

\bibliography{STAbib}

\end{document}